\makeatletter
\let\H@refstepcounter\refstepcounter
\makeatother

\documentclass[sigconf,natbib=true]{acmart}

\usepackage{multirow} 
\usepackage{multicol}
\usepackage{amsmath}
\usepackage{adjustbox}

\usepackage{algorithm}
\usepackage{algorithmic}

\AtBeginDocument{%
  \providecommand\BibTeX{{%
    \normalfont B\kern-0.5em{\scshape i\kern-0.25em b}\kern-0.8em\TeX}}}

\acmPrice{15.00}
\acmISBN{978-1-4503-XXXX-X/18/06}

\begin{document}

\title{SCaLRec: Semantic Calibration for LLM-enabled Cloud–Device Sequential Recommendation}

\author{Ruiqi Zheng}
\email{R.Zheng@latrobe.edu.au}
\affiliation{%
  \institution{La Trobe University}
  \city{Melbourne}
  \country{Australia}
}

\author{Jinli Cao}
\email{J.Cao@latrobe.edu.au}
\affiliation{%
  \institution{La Trobe University}
  \city{Melbourne}
  \country{Australia}
}

\author{Jiao Yin}
\email{jiao.yin@vu.edu.au}
\affiliation{%
  \institution{Victoria University}
  \city{Melbourne}
  \country{Australia}
}

\author{Hongzhi Yin}
\email{h.yin1@uq.edu.au}
\affiliation{%
  \institution{The University of Queensland}
  \city{Brisbane}
  \country{Australia}
}

\begin{abstract}

Cloud–device collaborative recommendation partitions computation across the cloud and user devices: the cloud provides semantic user modeling, while the device leverages recent interactions and cloud semantic signals for privacy-preserving, responsive reranking. With large language models (LLMs) on the cloud, semantic user representations can improve sequential recommendation by capturing high-level intent. However, regenerating such representations via cloud LLM inference for every request is often infeasible at real-world scale. As a result, on-device reranking commonly reuses a cached cloud semantic user embedding across requests. We empirically identify a cloud semantic staleness effect: reused embeddings become less aligned with the user’s latest interactions, leading to measurable ranking degradation.

Most existing LLM-enabled cloud–device recommenders are typically designed around on-demand cloud semantics, either by assuming low-latency cloud LLM access or by regenerating semantic embeddings per request. When per-request regeneration is infeasible and cached semantics must be reused, two technical challenges arise: (1) deciding when cached cloud semantics remain useful for on-device reranking, and (2) maintaining ranking quality when the cloud LLM cannot be invoked and only cached semantics are available. To address this gap, we introduce the Semantic Calibration for LLM-enabled Cloud–Device Recommendation (SCaLRec). First, it estimates the reliability of cached semantics under the user’s latest interactions. Second, an on-device semantic calibration module is proposed to adjusts the cached semantic embedding on-device using up-to-date interaction evidence, without per-request cloud LLM involvement. Experiments on real-world datasets show that SCaLRec consistently improves recommendation performance over strong baselines when cloud semantics are reused and cloud LLM inference is unavailable per request.

\end{abstract}

\ccsdesc[500]{Information systems~Recommender systems}

\keywords{Recommender Systems, Cloud-Device Collaboration}

\maketitle
\section{Introduction}

In recent years, sequential recommendation \cite{xu2021long,quadrana2018sequence} has become more significant, and have achieved success for both industry and academia. Graph neural networks and recurrent neural networks have shown promising recommendation performance by extracting the structural information and modeling behavioral pattern from the user-item interactions \cite{luo2024collaborative,donkers2017sequential}. However, those methods mainly focus on the clicking history and neglect the potential textual information inherited in the dataset and the high-level user preference. Large language models (LLM) offer a different type of signal for recommendation \cite{wu2024survey,lin2025can,zhao2024recommender}. They can summarize a user’s intent at a semantic level, rather than relying only on sparse item interactions. For instance, RARec \cite{yu2024ra} aligns ID representations with an LLM through soft prompts, so the model can exploit language knowledge. TransRec \cite{lin2024bridging} uses multi-facet identifiers such as IDs, titles, and attributes for distinct item generation and ranking.

Due to the nature of sequential recommendation, users' interests are heavily relied on the latest interactions \cite{devooght2017long}. This makes structural signals essential in the recent sequence. In many applications, however, these recent interactions are highly sensitive and inappropriate to directly upload to the cloud. In order to both make use of the semantic and structural information, one possible solution is to deploy the LLM on the device side \cite{lin2024awq,wang2025less}, so that the on-device LLM can extract the semantic signals and a traditional sequential model can learn the structural information. However, directly moving the large models from the cloud to the device is infeasible, considering the limited computation ability and resources on the device side. Only a compressed LLM can be a potential tradeoff, leading to performance degradation \cite{long2025cloud}. 

Another solution is the emerging device–cloud pattern for LLM-enabled recommendation \cite{zhan2025device,lv2025collaboration,long2025cloud}. The cloud hosts the large model and produces a semantic user representation when it is available. The device hosts a compact sequential model that tracks the newest interactions and provides a structure signal for reranking \cite{long2025cloud}. In inference stage, the reranker fuses the two signals. The system works well when the semantic signal is fresh.

However, in practice, the cloud LLM cannot be reliably invoked for every request \cite{luo2025toward}. Millions of users will ask for high volume of recommendation request at the same time. Utilizing the LLM only for inference on the powerful cloud server would still be costly. Consequently, the semantic signal is often cached and reused across requests rather than refreshed on demand. Most LLM-enabled device–cloud recommenders treat the cloud semantic signal as fresh by default \cite{long2025cloud}. If the signal is suspected to be outdated, a common fallback \cite{lv2025collaboration} is to invoke the cloud LLM again and overwrite the cache.

\begin{figure}[t]
\centering
\includegraphics[width=0.25\textwidth]{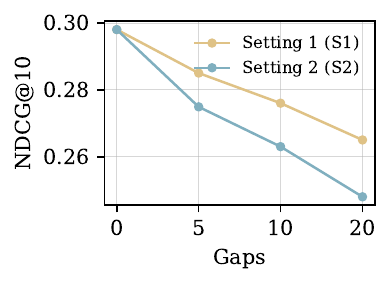} 
\vspace{-1.2em}
\caption{Cached cloud semantics becomes increasingly harmful as reuse grows. On Foursquare, reranking drops steadily with larger staleness gaps $g$ in both S1 (stale embedding only) and S2 (cached embedding and candidates).}
\label{fig:intro}
\vspace{-1.5em}
\end{figure}

We ask a basic question that is often overlooked: does semantic staleness actually happen in realistic recommendation, and how much can it hurt on-device reranking? We run a controlled diagnostic study on the Foursquare dataset \cite{yang2019revisiting}. At the current request time $t$, the device always uses the latest interactions to compute its structural signal. The difference is what the device reuses from a cached cloud snapshot generated at time $t-g$, where $g$ is the staleness gap measured in interaction steps. We consider two reuse patterns. In S1, we reuse the cached semantic user embedding but rerank a fresh candidate set produced at time $t$ (the latest top-$K$ candidates for on-device reranking), which isolates semantic–structure mismatch. In S2, we reuse both the semantic embedding and the candidate set from $t-g$, which matches an end-to-end cached pipeline when the cloud cannot be invoked to refresh either component. Figure~\ref{fig:intro} shows that ranking quality degrades as $g$ increases in both settings, and the drop is already visible at moderate gaps. This pattern suggests that semantic staleness can be a practical failure mode under intermittent cloud refresh, rather than a rare corner case.

This setting raises two challenges. First, bias estimation under missing fresh semantics: when the cloud LLM is unavailable and only cached semantics can be used, the system must infer how semantic mismatch distorts candidate ranking using only recent on-device interaction evidence and lightweight staleness cues. Second, bias-compensated robust fusion: the reranker should incorporate such compensation while preserving the standard semantic–structured fusion form, so that ranking quality degrades gracefully as mismatch grows, instead of abruptly collapsing by over-trusting stale semantics or discarding semantic evidence entirely.

To this end, we propose a novel calibration framework, namely Semantic Calibration for LLM-enabled Cloud–Device Recommendation (SCaLRec). To make the cached semantic signal usable under the cloud staleness effect, SCaLRec separates the problem into two lightweight modules on the device. It first proposes an on-device reliability estimator. The inputs for the estimator are cached semantic user embedding current and previous structural embeddings. The outputs are predictions on whether the semantic channel is still trustworthy for the current request. A straightforward approach is to use a learnable gate to control the semantic score when the estimator reports low reliability, and it relies more on the structural signal. However, simple gating would discard semantic evidence even when parts of it remain informative. In our experiments, this naive gating leads to a noticeable loss of the semantic benefit, especially at moderate staleness gaps.

The second module is a semantic calibrator. It tries to recover useful semantic evidence from a cached cloud user embedding, instead of simply down-weighting it. The calibrator predicts an embedding-level residual update and applies it to the cached semantic embedding on device. We train this correction with knowledge distillation. In offline training, we can access a fresh semantic embedding for the same user state and treat it as a teacher signal. The teacher reranker uses the fresh semantic embedding and produces soft scores over the candidate set. The student reranker uses the calibrated cached embedding and produces its own score distribution. The calibrator is optimized to make the student match the teacher, so the calibrated semantics induces a similar ranking behavior to the fresh semantics. We do not aim to perfectly reconstruct the fresh embedding itself, but to reconstruct its effect on reranking. At inference time, the calibrator runs fully on device and only needs the cached semantic embedding, the current structure signal from recent interactions, and lightweight staleness cues. The reliability estimator further controls the correction strength, so the calibrator makes stronger updates when mismatch is likely and stays conservative when the cached semantics remains consistent.

The contributions are summarized as follows: 
\begin{itemize}
    \item We identify semantic staleness in LLM-enabled device–cloud sequential reranking, and we formalize it as a semantic–structure mismatch under intermittent refresh. We further introduce a controlled evaluation protocol that separates embedding staleness from end-to-end cached pipeline effects.
    \item We propose SCaLRec, a lightweight on-device calibration framework that consists of a reliability estimator and a semantic calibrator. The estimator predicts when cached semantics is risky, and the calibrator recovers useful semantic evidence by learning an embedding-level correction through knowledge distillation from fresh semantics available in offline training.
    \item We conduct comprehensive experiments on two real-world sequential recommendation datasets under varying staleness gaps and availability conditions. Results show consistent improvements over strong device–cloud and on-device baselines, with ablation and model agnostic study.
\end{itemize}

\section{Related Work}

\subsection{Device-Cloud Recommendation}
\label{sec:rw:dc}

Device-cloud recommendation splits the pipeline across a cloud service and user devices for privacy and efficiency. MetaController determine whether use the device model only or rely on the cloud by a dynamic collaboration pipeline. DualRec \cite{zhang2025dualrec} jointly trains lightweight device models and a larger cloud model via bidirectional knowledge transfer without sharing raw interactions. IntellectReq \cite{lv2024intelligent} and DUET \cite{lv2023duet} generate personalized parameters on the cloud side to enhance the device model. CDA4Rec \cite{long2025cloud} uses cloud-hosted LLMs for task decomposition and planning to instruct the local model. LSC4Rec \cite{lv2025collaboration} emphasizes the system trade-off between cloud semantic strength and device efficiency.

\subsection{Interest Drift, Stale Embeddings, and Cloud Semantic Staleness}
\label{sec:rw:drift-stale}

User interest drift is a classic topic in recommender systems, and many temporal or sequential models introduce time-aware components to track evolving preferences \cite{lin2025towards,hyun2022beyond,chen2021multi}.
Concept drift is also studied as a broader learning problem, and it motivates online adaptation and evaluation protocols under non-stationary data \cite{gama2014survey}.
Embedding staleness is often discussed at the representation level, and it refers to outdated item embeddings caused by delayed retraining, or delayed indexing \cite{zeng2024accelerating}. A related line of work proposes corrector-style networks to reduce the damage of stale dense embeddings in retrieval systems \cite{monath2024fresh}.

Unlike prior work on interest drift or stale embeddings that treats staleness as a single-channel degradation, we study a device-cloud reranking setting where stale cloud semantics could possibly remain individually reasonable yet become harmful when fused with fresh on-device structural signals, yielding a structured fusion bias induced by cross-channel freshness mismatch.

\section{Preliminaries}

\textbf{Sequential Recommendation}
We denote $U$ as the set of users and $I$ as the set of items. For each user $u\in U$, we observe a chronologically ordered interaction sequence
$X_u=[i_1,\ldots,i_{T_u}]$ with $i_t\in I$. At time step $t$, the model takes the prefix
$X_{u,t}=[i_1,\ldots,i_t]$ and aims to predict the next item $i_{t+1}$.
Each item $i\in I$ is associated with an embedding vector $e_i\in\mathbb{R}^d$.

\textbf{Cloud-Device Recommendation Framework} We consider a device-cloud sequential recommendation pipeline with two complementary signals.
On the device, a lightweight sequential encoder extracts a real-time \emph{structure signal} from recent interactions:
\begin{equation}
e^{str}_{u,t}=f_{\text{dev}}(X_{u,t})\in\mathbb{R}^d .
\end{equation}
On the cloud, a large model (e.g., an LLM) produces a high-level \emph{semantic signal} by modeling user intent from a compact summary.
Let $a_{u,t}$ denote an intent summary constructed from interaction evidence, and optionally a natural-language request $q_{u,t}$ from user $u$ when available:
\begin{equation}
a_{u,t}=f_{\text{abs}}(X_{u,t}, q_{u,t}).
\end{equation}
The cloud semantic signal is then instantiated as a semantic user embedding
\begin{equation}
e^{sem}_{u,t}=f_{\text{cloud}}(a_{u,t})\in\mathbb{R}^d .
\end{equation}
In practice, the system reuses a cached semantic embedding generated at an earlier time step $\tau<t$, denoted by $e^{sem}_{u,\tau}$.

\textbf{
Difference from Interest Shift and Embedding Staleness.}
Interest shift concerns how user preference evolves over time, whereas cloud semantic staleness is an architecture-level mismatch caused by asynchronous updates of cloud semantic and on-device structural signals when cloud semantics cannot be refreshed on demand. Notably, staleness can arise even without major preference drift, because the relevance of high-level semantics versus recent interactions may change with situational context (e.g., location or activity), making cached semantics misaligned. It also differs from embedding staleness, which typically stems from delayed retraining or model-version lag. Our study isolates this mismatch and mitigates its impact during on-device reranking.

\textbf{Problem Formulation}
We focus on the reranking stage under cached cloud semantics. Given the cached semantic signal $e^{sem}_{u,\tau}$ and the on-device structure signal $e^{str}_{u,t}$, the backbone computes a semantic-structured fused score for each item $i\in I$ as
\begin{equation}
\text{score}_{u,t}(i)
=
\beta\,\langle e^{sem}_{u,\tau}, e_i\rangle
+
(1-\beta)\,\langle e^{str}_{u,t}, e_i\rangle,
\label{eq:rerank}
\end{equation}
where $\langle\cdot,\cdot\rangle$ is the dot product and $\beta\in[0,1]$ is a fusion weight set as in the backbone, and the device returns the top-$K$ items.
Our goal is to improve ranking quality under intermittent cloud availability, i.e., to mitigate the mismatch-induced bias caused by semantic staleness and to ensure that ranking performance degrades gracefully as the staleness gap $g$ increases, without relying on per-request cloud LLM invocation.

\begin{figure}[t]
\centering
\includegraphics[width=0.5\textwidth]{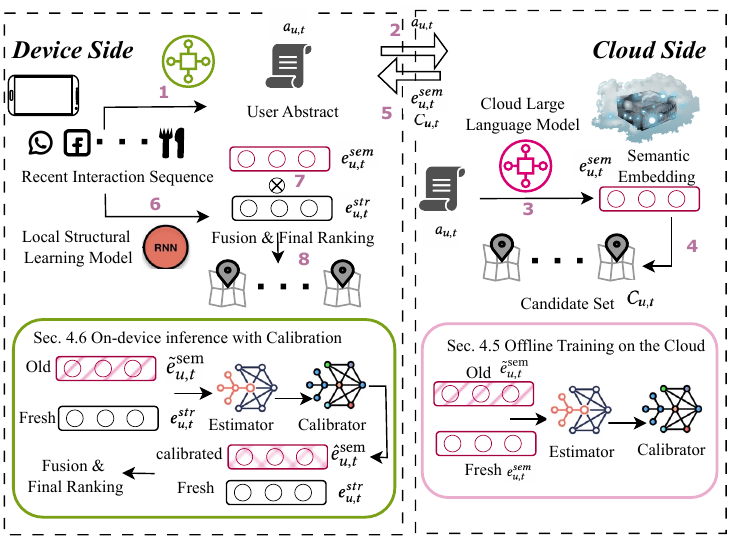} 
\vspace{-2.5em}
\caption{Overview of the proposed SCaLRec.}
\label{fig:main}
\vspace{-1.5em}
\end{figure}

\section{SCaLRec}
\label{sec:method}

We study a device-cloud sequential reranking setting \cite{zhan2025device,lv2025collaboration,long2025cloud}.
The cloud provides a semantic signal from an LLM.
The device provides a structure signal from recent interactions, and ranks a candidate set under a fixed dot-product interface as shown in Figure~\ref{fig:main}.
The cloud cannot be invoked on demand for every request.
So the device often reuses a cached cloud snapshot.
This is the place where semantic staleness shows up.

\subsection{Backbone: Device-Cloud Sequential Reranking}
\label{sec:method:backbone}

User $u$ has a recent interaction sequence at step $t$,
denoted by $X_{u,t} = [i_{t-L+1}, \dots, i_t]$ with length $L$.
Some datasets also provide a natural-language query $q_{u,t}$.
If $q_{u,t}$ is not available, we set $q_{u,t}=\emptyset$ and rely on $X_{u,t}$.

\textit{Item embedding initialization.} Each item $i \in \mathcal{I}$ has an embedding $e_i \in \mathbb{R}^d$. We collect them as an embedding matrix $E \in \mathbb{R}^{|\mathcal{I}|\times d}$.
The cloud maintains a large language model $\Theta_{\ell}$.
Each item has textual metadata $\texttt{text}(i)$, such as title and tags.
We compute a high-dimensional representation by averaging token states,
denoted by $h_i = \texttt{AVG}(\Theta_{\ell}(\texttt{text}(i)))$.
We then apply an offline dimensionality reduction step $\texttt{PCA}(\cdot)$.
It maps $h_i$ into the target space $e_i = \texttt{PCA}\!\left(h_i\right)$ where $e_i \in \mathbb{R}^d$.

\textit{Device-to-cloud abstraction.}
The device builds a compact textual abstract from local signals.
The abstract summarizes the request intent and recent behaviors.
It uses $q_{u,t}$ and $X_{u,t}$.
We denote the abstract by $a_{u,t}$.
A simple form is a concatenation of a behavioral summary and a sanitized query.
We write it as
\begin{equation}
a_{u,t} = \texttt{AbsGen}(q_{u,t}, X_{u,t}),
\label{eq:abstract}
\end{equation}
where $\texttt{AbsGen}(\cdot)$ is a lightweight on-device module. In our experiments, we instantiate the cloud model $\Theta_{\ell}$ as LLaMA-3.1-8B \cite{touvron2023llama} and the on-device model $\Theta_s$ as LLaMA-3.2-1B for abstract generation. 

\textit{Cloud semantic user modeling and candidate retrieval.}
The cloud encodes $a_{u,t}$ into a semantic user embedding $e^{\text{sem}}_{u,t} \in \mathbb{R}^d$:
\begin{equation}
e^{\text{sem}}_{u,t} = f_{\text{sem}}(a_{u,t}; \Theta_{\ell}),
\label{eq:sem-user}
\end{equation}
where $f_{\text{sem}}(\cdot)$ is the semantic user modeling function on the cloud.
The cloud then retrieves a candidate set of size $N$ from the item lists.
We denote the candidate set by $C_{u,t} \subset \mathcal{I}$:
\begin{equation}
C_{u,t} = \texttt{Top-}N\Big(\{\langle e^{\text{sem}}_{u,t}, e_i \rangle\}_{i \in \mathcal{I}}\Big).
\label{eq:candidate}
\end{equation}

\textit{On-device structured modeling.}
The device runs a lightweight sequential encoder over $X_{u,t}$.
It outputs a structured user embedding: 
\begin{equation}
e^{\text{str}}_{u,t} = f_{\text{enc}}(X_{u,t}; \phi),
\label{eq:str-user}
\end{equation}
where $\phi$ are on-device model parameters.
This encoder can be SASRec \cite{kang2018self} or SURGE \cite{chang2021sequential} in our instantiations. In Section~\ref{sec:rq4}, we discuss the selection of the local structural signal learning models, and an ablation study shows that SCaLRec does not depend on one specific model. The proposed method can constantly outperform the base recommendation model through the combination of structural and semantical signals, with the help of semantic calibration.

\subsection{Caching and Cloud Semantic Staleness}
\label{sec:method:staleness}

\subsubsection{Cached cloud snapshot.}
The cloud is not invoked for every request in a deployed device-cloud pipeline.
A common implementation reuses the most recent cloud output for multiple subsequent requests.
We treat this reuse as a given system constraint.
We do not model or optimize any refresh policy.

For user $u$ at step $t$, let $\tau_{u,t}$ denote the time index of the most recent cloud snapshot available \emph{before} $t$.
We denote the cached semantic user embedding by $\tilde e^{\mathrm{sem}}_{u,t} \in \mathbb{R}^d$ and the cached candidate set by $\tilde C_{u,t}\subset\mathcal{I}$ with $|\tilde C_{u,t}|=N$.
The device also caches the structure embedding computed at the snapshot time, denoted by $\tilde e^{\mathrm{str}}_{u,t}\in\mathbb{R}^d$. Therefore, the cached snapshot can be written as:
\begin{equation}
\mathcal{Z}_{u,t} = (\tau_{u,t}, \tilde e^{\mathrm{sem}}_{u,t}, \tilde C_{u,t}).
\label{eq:cache_state_v2}
\end{equation}
The device always computes the up-to-date structure embedding $e^{\mathrm{str}}_{u,t}$ from the latest sequence $X_{u,t}$ by Eq.~\eqref{eq:str-user}.

\subsubsection{Architecture-induced mismatch.}
Cloud semantic staleness refers to a mismatch between two signals used in on-device reranking.
The structure signal $e^{\mathrm{str}}_{u,t}$ is updated at every request.
The semantic signal is reused from $\tau_{u,t}$ as $\tilde e^{\mathrm{sem}}_{u,t}$.
These two signals can become asynchronous in relevance.
The fused score in Eq.~\eqref{eq:rerank} can then be biased by over-trusting the cached semantic channel.

For offline analysis, we define an unobserved semantic error $\epsilon_{u,t} = \left\| e^{\mathrm{sem}}_{u,t} - \tilde e^{\mathrm{sem}}_{u,t} \right\|_2 .$
$\epsilon_{u,t}$ is not available when the cloud is not called.
We only use it for training supervision and diagnostics.

As for the staleness cues on the device, the device can compute lightweight cues to assess mismatch risk.
We use a time-gap feature $\delta_{u,t} = t - \tau_{u,t}$, and compute the changes of structural signal by the previous structure embedding $e^{\mathrm{str}}_{u,\tau_{u,t}}$ stored at the snapshot time. The changes can be recoreded as $\Delta^{\mathrm{str}}_{u,t} = \left\| e^{\mathrm{str}}_{u,t} - e^{\mathrm{str}}_{u,\tau_{u,t}} \right\|_2$.

We further use a semantic-structure agreement signal on the cached candidate set $\tilde C_{u,t}$.
We compute two score lists on $\tilde C_{u,t}$,
$s^{\mathrm{sem}}_{u,t}(i)=\langle \tilde e^{\mathrm{sem}}_{u,t}, e_i\rangle$
and
$s^{\mathrm{str}}_{u,t}(i)=\langle e^{\mathrm{str}}_{u,t}, e_i\rangle$.
We then measure their agreement by the Spearman rank correlation:
\begin{equation}
\omega_{u,t}
=
\rho_{\mathrm{sp}}\!\Big(
\mathrm{rank}\big(\{s^{\mathrm{sem}}_{u,t}(i)\}_{i\in \tilde C_{u,t}}\big),
\mathrm{rank}\big(\{s^{\mathrm{str}}_{u,t}(i)\}_{i\in \tilde C_{u,t}}\big)
\Big),
\label{eq:omega_spearman}
\end{equation}
where $\rho_{\mathrm{sp}}(\cdot,\cdot)\in[-1,1]$ is the Spearman correlation and $\mathrm{rank}(\cdot)$ maps scores to ranks within $\tilde C_{u,t}$.
In case of ties, we use average ranks. We care about how the two channels order the candidates, not only their geometric distance in the embedding space.
We therefore use a rank-based agreement signal, which is directly aligned with reranking quality.
We collect these cues into
$c_{u,t} = [\delta_{u,t}, \Delta^{\mathrm{str}}_{u,t}, \omega_{u,t}, \mathrm{stats}_{u,t}]$.
$\mathrm{stats}_{u,t}$ includes simple device-side statistics such as sequence length and interaction density.
Note that $\omega_{u,t}$ is computed on the reranking candidate set $\tilde C_{u,t}$, so the cost is bounded by the candidate size rather than the full item corpus.

\subsubsection{Controlled staleness protocol.}
We simulate semantic staleness by reusing an earlier semantic snapshot.
For a user at step $t$, we replace the semantic embedding with one generated at $t-g$.
$g$ is the staleness gap measured in interaction steps.
We keep the device encoder and the sequence input at $t$ unchanged.
This isolates the mismatch introduced by caching, rather than changing the on-device evidence.

In many device-cloud pipelines, candidate retrieval is driven by the semantic embedding.
A stale semantic snapshot then implies a stale retrieval snapshot.
We therefore consider two settings.

S1 is a diagnostic setting.
We keep the candidate set fixed to the fresh one at step $t$.
We only reuse a stale semantic embedding:
\begin{equation*}
\textbf{S1:}\quad
\hat C_{u,t} = C_{u,t},\qquad
\hat e^{\mathrm{sem}}_{u,t} = e^{\mathrm{sem}}_{u,t-g}.
\label{eq:s1_v2}
\end{equation*}

S2 reflects an end-to-end cached pipeline.
Both the semantic embedding and the candidate set come from the same stale snapshot:
\begin{equation*}
\textbf{S2:}\quad
\hat C_{u,t} = C_{u,t-g},\qquad
\hat e^{\mathrm{sem}}_{u,t} = e^{\mathrm{sem}}_{u,t-g}.
\label{eq:s2_v2}
\end{equation*}

In both settings, the device still computes $e^{\mathrm{str}}_{u,t}$ from the latest interactions.
The reranker uses the same fusion form as Eq.~\eqref{eq:rerank}.
Only the simulated semantic channel and the simulated retrieval snapshot change.

\subsection{Semantic Reliability Estimation}
\label{sec:method:estimator}

The device needs a signal to describe how risky the cached semantic channel is.
We do not treat it as a refresh decision.Instead, we consider it as a reliability score for the cached snapshot.
The score only controls how much correction we apply, and it does not trigger any cloud call. Based on the device-observable cue vector $c_{u,t}$, which includes $\delta_{u,t}$, $\Delta^{\mathrm{str}}_{u,t}$, and $\omega_{u,t}$.
We use a lightweight estimator $q_{\psi}(\cdot)$:
\begin{equation}
r_{u,t} = q_{\psi}(c_{u,t}) \in [0,1].
\label{eq:reliability}
\end{equation}
$r_{u,t}$ close to $1$ means that the cached semantic signal looks consistent with the latest evidence on the device, while $r_{u,t}$ close to $0$ means that it becomes unreliable.
This is an affordable check, and only runs on the device, without the cloud dependency. We convert $r_{u,t}$ to a correction scale $\alpha_{u,t} = 1 - r_{u,t}$ for later usage in semantic compensation. $\alpha_{u,t}$ will be used to scale the correction radius in Sec.~\ref{sec:method:calibration}, and the training procedure for $q_{\psi}$ is described in Sec.~\ref{sec:method:training}.

We use $r_{u,t}$ to scale the correction magnitude. When the snapshot looks consistent, the update becomes close to an identity mapping. It also allows larger corrections when the mismatch risk is high. Unlike a total end-to-end calibrator, we decouple the process into an estimator and a calibrator guided by the estimator. In practice, a calibrator alone may introduce unnecessary changes at small gaps. The reliability score provides a simple knob to keep the update close to identity when mismatch evidence is weak.


\subsection{Bias-Compensated Semantic Calibration}
\label{sec:method:calibration}

The reliability score in Eq.~\eqref{eq:reliability} tells us how risky the cached semantic channel is.
It does not tell us how to fix it.
We learn a calibrator to reduce the score bias caused by semantic mismatch on the device side.
We do not introduce any refresh rule related to the cloud.

\paragraph{Calibrator input and output.}
We use a lightweight calibrator $g_{\phi}(\cdot)$.
It takes the cached semantic embedding and the latest structure embedding:
\begin{equation}
v_{u,t} = g_{\phi}\big(\tilde e^{\mathrm{sem}}_{u,t},\ e^{\mathrm{str}}_{u,t}\big) \in \mathbb{R}^d.
\label{eq:calibrator_raw}
\end{equation}
We treat $v_{u,t}$ as a correction proposal. In order to scale its magnitude to satisfy a bounded update, we set a maximum correction radius $\gamma_{\max}>0$, and scale it with $\alpha_{u,t}=1-r_{u,t}$ from Sec.~\ref{sec:method:estimator}.
We form a bounded correction by scaling the proposal vector:
\begin{equation}
\begin{aligned}
\Delta e_{u,t} &= \lambda_{u,t}\, v_{u,t},\\
\lambda_{u,t} &=
\begin{cases}
1, & \|v_{u,t}\|_2 \le \gamma_{\max}\cdot \alpha_{u,t},\\[4pt]
\dfrac{\gamma_{\max}\cdot \alpha_{u,t}}{\|v_{u,t}\|_2 + \varepsilon},
& \|v_{u,t}\|_2 > \gamma_{\max}\cdot \alpha_{u,t}.
\end{cases}
\end{aligned}
\label{eq:delta_e_bounded}
\end{equation}
$\varepsilon$ is a small constant.
This guarantees $\|\Delta e_{u,t}\|_2 \le \gamma_{\max}\cdot \alpha_{u,t}$.
We then produce the corrected semantic embedding as
$\hat e^{\mathrm{sem}}_{u,t} = \tilde e^{\mathrm{sem}}_{u,t} + \Delta e_{u,t}$.

\paragraph{Reranking with the corrected semantic channel.}
We keep the same fusion interface as Eq.~\eqref{eq:rerank}.
We only replace the semantic embedding by $\hat e^{\mathrm{sem}}_{u,t}$:
\begin{equation}
\hat s_{u,t}(i) =
\beta \cdot \langle \hat e^{\mathrm{sem}}_{u,t}, e_i\rangle
+
(1-\beta)\cdot \langle e^{\mathrm{str}}_{u,t}, e_i\rangle,
\qquad i \in \tilde C_{u,t}.
\label{eq:rerank_corrected}
\end{equation}

This update only changes the semantic embedding used in reranking.
It does not update model parameters at test time, which differs from test-time parameter correction methods such as CoCorrRec.
We do not add a separate cloud-calling decision.
The calibrator outputs an embedding-level correction applied within the same dot-product reranker.
\begin{algorithm}[t]
\caption{On-device inference with reliability-guided semantic calibration}
\label{alg:inference}
\begin{algorithmic}[1]
\REQUIRE Cached snapshot $(\tilde e^{\mathrm{sem}}_{u,t},\tilde C_{u,t},\tau_{u,t},\tilde e^{\mathrm{str}}_{u,t})$; recent interactions $X_{u,t}$; item embeddings $\{e_i\}$.
\ENSURE Top-$K$ recommendation list.
\STATE Compute latest structure embedding $e^{\mathrm{str}}_{u,t}=f_{\mathrm{enc}}(X_{u,t})$.
\STATE Compute gap $\delta_{u,t}=t-\tau_{u,t}$ and drift cue $\Delta^{\mathrm{str}}_{u,t}=\|e^{\mathrm{str}}_{u,t}-\tilde e^{\mathrm{str}}_{u,t}\|_2$.
\STATE For each $i\in \tilde C_{u,t}$, compute $s^{\mathrm{sem}}_{u,t}(i)=\langle \tilde e^{\mathrm{sem}}_{u,t}, e_i\rangle$ and $s^{\mathrm{str}}_{u,t}(i)=\langle e^{\mathrm{str}}_{u,t}, e_i\rangle$.
\STATE Compute $\omega_{u,t}$ by Spearman correlation on $\tilde C_{u,t}$ (or top-$L$).
\STATE Form cues $c_{u,t}=[\delta_{u,t},\Delta^{\mathrm{str}}_{u,t},\omega_{u,t},\mathrm{stats}_{u,t}]$.
\STATE Predict reliability $r_{u,t}=q_{\psi}(c_{u,t})$ and set $\alpha_{u,t}=1-r_{u,t}$.
\STATE Predict proposal vector $v_{u,t}=g_{\phi}(\tilde e^{\mathrm{sem}}_{u,t}, e^{\mathrm{str}}_{u,t})$.
\STATE Compute bounded correction $\Delta e_{u,t}$ by Eq.~\eqref{eq:delta_e_bounded}.
\STATE Correct semantic embedding $\hat e^{\mathrm{sem}}_{u,t}=\tilde e^{\mathrm{sem}}_{u,t}+\Delta e_{u,t}$.
\STATE For each $i\in \tilde C_{u,t}$, compute corrected reranking score
$\hat s_{u,t}(i)=\beta\langle \hat e^{\mathrm{sem}}_{u,t}, e_i\rangle+(1-\beta)\langle e^{\mathrm{str}}_{u,t}, e_i\rangle$.
\STATE Return top-$K$ items in $\tilde C_{u,t}$ by $\hat s_{u,t}(i)$.
\end{algorithmic}
\end{algorithm}

\subsection{Offline Training of Estimator and Calibrator}
\label{sec:method:training}

We train the reliability estimator $q_{\psi}$ and the calibrator $g_{\phi}$ offline.
This follows the common practice in sequential recommendation research.
After training, both modules are deployed on the device for inference.
The inference stage does not upload raw interaction sequences to the cloud. 

We sample a user $u$ and a target step $t$ from the training log, which are desensitized historical interactions or public interaction shared with the cloud.
We then sample a staleness gap $g$ from a predefined set $\mathcal{G} = \{5, 10, 20 \}$.
We construct a stale semantic snapshot from step $t-g$.
We denote the stale semantic embedding by
$\tilde e^{\mathrm{sem}}_{u,t} = e^{\mathrm{sem}}_{u,t-g}$. The reranking candidate set is denoted by $\tilde C_{u,t}$.
We consider two standard constructions that match S1 and S2.

In both settings, the device-side encoder still uses the latest interactions at $t$ as $e^{\mathrm{str}}_{u,t} = f_{\mathrm{enc}}(X_{u,t})$.
We also store the structure embedding at the snapshot step as a cache proxy as 
$\tilde e^{\mathrm{str}}_{u,t} = f_{\mathrm{enc}}(X_{u,t-g})$. We compute a fresh semantic embedding as a teacher for training only:
$e^{\mathrm{sem}}_{u,t}$.
It is available offline and not required at inference time.
Teacher and stale score lists on $\tilde C_{u,t}$ are formed as below:
\begin{equation}
z_{u,t}(i)=\langle e^{\mathrm{sem}}_{u,t}, e_i\rangle,\qquad
\tilde z_{u,t}(i)=\langle \tilde e^{\mathrm{sem}}_{u,t}, e_i\rangle,\qquad
i\in \tilde C_{u,t}.
\label{eq:teacher_stale_scores}
\end{equation}
We convert them into distributions with temperature $T$:
\begin{equation}
p_{u,t}=\mathrm{softmax}(z_{u,t}/T),\qquad
\tilde p_{u,t}=\mathrm{softmax}(\tilde z_{u,t}/T).
\label{eq:teacher_stale_dist}
\end{equation}

\paragraph{Reliability supervision.}
We use the cue vector $c_{u,t}$ from Section~\ref{sec:method:staleness} and predict $r_{u,t}=q_{\psi}(c_{u,t})$.
We define a mismatch severity score by KL divergence:
\begin{equation}
d_{u,t}=\mathrm{KL}\big(p_{u,t}\ \|\ \tilde p_{u,t}\big).
\label{eq:severity_kl}
\end{equation}
We map it to a bounded target in $[0,1]$:
\begin{equation}
y_{u,t}=\exp(-d_{u,t}).
\label{eq:reliability_target}
\end{equation}
The estimator loss is a regression loss:
\begin{equation}
\mathcal{L}_{\mathrm{rel}}=\big\|q_{\psi}(c_{u,t})-y_{u,t}\big\|_2^2.
\label{eq:loss_rel_v2}
\end{equation}

\paragraph{Calibration distillation.}
Given $r_{u,t}$, we compute $\alpha_{u,t}=1-r_{u,t}$.
We compute the bounded correction in Eq.~\eqref{eq:delta_e_bounded} and obtain the corrected embedding
$\hat e^{\mathrm{sem}}_{u,t}=\tilde e^{\mathrm{sem}}_{u,t}+\Delta e_{u,t}$.
We then form the corrected semantic score list:
\begin{equation}
\hat z_{u,t}(i)=\langle \hat e^{\mathrm{sem}}_{u,t}, e_i\rangle,\qquad
\hat p_{u,t}=\mathrm{softmax}(\hat z_{u,t}/T),\qquad
i\in \tilde C_{u,t}.
\label{eq:student_scores}
\end{equation}

We distill the teacher ranking behavior with a KL loss and add a small $\ell_2$ regularizer on the proposal vector $v_{u,t}$. We train $q_{\psi}$ and $g_{\phi}$ jointly with the following objective:
\begin{equation}
\begin{aligned}
\mathcal{L}_{\mathrm{KD}} &= \mathrm{KL}\big(p_{u,t}\ \|\ \hat p_{u,t}\big),\\
\mathcal{L}_{\mathrm{reg}} &= \|v_{u,t}\|_2^2,\\
\mathcal{L} &=
\mathcal{L}_{\mathrm{KD}}
+
\lambda_{\mathrm{rel}}\mathcal{L}_{\mathrm{rel}}
+
\lambda_{\mathrm{reg}}\mathcal{L}_{\mathrm{reg}}.
\end{aligned}
\label{eq:loss_all_v2}
\end{equation}

\subsection{On-device Inference}
\label{sec:method:inference}

We now describe the on-device inference procedure.
The device receives the cached snapshot $\mathcal{Z}_{u,t}$ from the last successful cloud response.
It includes the cached semantic embedding $\tilde e^{\mathrm{sem}}_{u,t}$ and the reranking candidate set $\tilde C_{u,t}$.
The device also has the latest interaction prefix $X_{u,t}$.

At each request, the device encodes $X_{u,t}$ and obtains a fresh structure embedding $e^{\mathrm{str}}_{u,t}$.
It then computes lightweight staleness cues.
These cues include the gap $\delta_{u,t}$, a structure drift magnitude, and a semantic-structure agreement score on $\tilde C_{u,t}$.
We use them to predict a reliability score $r_{u,t}$.
The score does not trigger any cloud call.
It only controls how strong the correction can be.

Next, the calibrator proposes an embedding-level update for the cached semantic embedding.
We apply a bounded scaling so the update stays within a controlled radius.
This keeps the behavior close to identity when mismatch evidence is weak.
We then rerank items in $\tilde C_{u,t}$ using the same dot-product fusion form as the backbone.
Algorithm~\ref{alg:inference} summarizes the full inference steps.

\subsection{Complexity and Practical Notes}
\label{sec:method:complexity}

The estimator uses cues computed on the reranking candidate set $\tilde C_{u,t}$.
The most expensive cue is the Spearman correlation in Eq.~\eqref{eq:omega_spearman}.
It requires sorting the semantic and structure scores.
The cost is $O(|\tilde C_{u,t}|\log |\tilde C_{u,t}|)$.
This cost is bounded by the reranking candidate size. In practice, we compute $\omega_{u,t}$ on the top-$L$ candidates under the semantic scores.
$L$ is a small constant such as $100$ or $200$.
This keeps the cost stable on device.
All other computations are linear in $|\tilde C_{u,t}|$.

\section{Experiments}
\label{sec:exp}
To validate the effectiveness of SCaLRec, we conduct various experiments to address the following five research quesstions (RQs).

\textbf{RQ1}: Does SCaLRec consistently improve reranking performance over strong baselines under intermittent cloud availability?

\textbf{RQ2}: How do different methods degrade as the staleness gap increases under (S1) controlled reranking-only staleness and (S2) realistic pipeline staleness?

\textbf{RQ3}: How sensitive is SCaLRec to key hyperparameters?

\textbf{RQ4}: Is SCaLRec model-agnostic, i.e., can it be plugged into different device-cloud backbones with minimal changes?

\textbf{RQ5}: Where do the gains come from? (Ablations and common staleness mitigation tricks.)


\subsection{Experimental Settings}

\subsubsection{Datasets}
\label{sec:exp_datasets}
We evaluate on the two widely used datasets to assess SCaLRec, covering both conversational recommendation and large-scale implicit feedback. \textbf{ReDial} is movie recommendation with conversational context and item recommendation signals. \textbf{Foursquare} is Point-of-Next-Interest recommendation task with spatio-temporal check-in sequences. Table~\ref{tab:datasets} reports the statistics after preprocessing.

\subsubsection{Preprocessing and Data Split}
\label{sec:exp_split}
For each user, interactions are sorted by timestamp.
We adopt a leave-one-out split, where 
the last interaction is used for testing, and the second last is for validation, and the remaining interactions are for training.
We truncate each user sequence to a maximum length of 200.
Unless otherwise stated, we use next-item prediction evaluation and report ranking metrics at $K{=}10$.

\subsubsection{Metrics}
\label{sec:exp_metrics}

We use HR@$K$ and NDCG@$K$ to assess ranking quality. HR@$K$ indicates whether the ground-truth item is included in the top-$K$ recommendations, and NDCG@$K$ places more emphasis on ranking it closer to the top.

\subsubsection{Optimization and hyperparameters.}
Adam optimizer is used with batch size 64, dropout 0.2, and learning rate $2\times 10^{-4}$. We set the embedding dimension to $d{=}1024$. Candidate set size $N{=}1000$ by default. The reliability estimator $q_\psi(\cdot)$ is a 2-layer MLP with hidden size $64$ and ReLU activation.
The calibrator $g_\phi(\cdot)$ is also a lightweight 2-layer MLP with hidden size $256$ and ReLU activation. The Spearman agreement $\omega_{u,t}$ is computed on the cached reranking set $\tilde C_{u,t}$.
To bound the on-device cost, we compute it on the top-$L$ items and set $L{=}200$ unless specified.
We train for at most $20$ epochs with early stopping and select the best checkpoint by validation NDCG@10. For SCaLRec-specific hyperparameters, $\gamma_{\max}$, $T$, $\lambda_{\mathrm{rel}}$, $\lambda_{\mathrm{reg}}$ are set as 0.2, 4, 1, $10^{-3}$ correspondingly.

\begin{table}[t]
\centering
\begin{tabular}{c|c|c|c}
\toprule
      & \#User & \#Item & \#Interactions \\
\midrule
\midrule
ReDial & 1,482 & 33,834 & 300,401 \\
Foursquare & 1,083 & 38,333 & 227,428 \\
\bottomrule
\end{tabular}%

\caption{Dataset statistics after preprocessing.}
\label{tab:datasets}
\vspace{-3em}

\end{table}

\subsection{Intermittent Cloud Simulation and Staleness Protocol}
\label{sec:exp_staleness}
A key practical constraint is that the cloud semantic signal cannot be refreshed on demand.
We simulate intermittent cloud availability by allowing a cloud refresh only once every $g$ requests.
At a refresh step $\tau$, the cloud produces a semantic user embedding and a candidate set.
For requests $t\in(\tau,\tau{+}g]$, the system reuses cached cloud outputs.

We evaluate two complementary staleness settings:

\textbf{(S1) Reranking-only staleness (controlled).}
We keep the candidate set $C_{u,t}$ fixed and only replace the semantic embedding at time $t$
with a cached embedding generated at an earlier time $\tau$.
This isolates the \emph{semantic mismatch} effect on reranking.

\textbf{(S2) Pipeline staleness (realistic).}
Both the semantic embedding and the candidate set are cached from $\tau$ and reused for $t\in(\tau,\tau{+}g]$.
This reflects the realistic case where cloud retrieval is also unavailable during the gap.

\begin{table*}[htbp]
\renewcommand{\arraystretch}{0.3}
  \centering
  \caption{The recommendation performance of the proposed SCaLRec and competitive baselines on two real-world datasets.}

\begin{tabular}{c|c|cccc|cccc}
\toprule
\multirow{2}[2]{*}{Category} & \multirow{2}[2]{*}{Method} & \multicolumn{4}{c|}{Foursquare} & \multicolumn{4}{c}{ReDial} \\
      &       & NDCG@5 & HR@5  & NDCG@10 & HR@10 & NDCG@5 & HR@5  & NDCG@10 & HR@10 \\
\midrule
\midrule
\multirow{5}[2]{*}{Cloud Rec } & LightGCN & 0.1795  & 0.2411  & 0.2109  & 0.3578  & 0.0068  & 0.0109  & 0.0096  & 0.0211  \\
      & SURGE & 0.1898  & 0.2578  & 0.2318  & 0.3752  & 0.0074  & 0.0115  & 0.0094  & 0.0227  \\
      & SASRec  & 0.1867  & 0.2675  & 0.2273  & 0.3596  & 0.0075  & 0.0116  & 0.0109  & 0.0229  \\
      & RARec & 0.2293  & 0.3190  & 0.2433  & 0.4097  & 0.0162  & 0.0230  & 0.0230  & 0.0378  \\
      & TransRec & 0.2087  & 0.2813  & 0.2284  & 0.4107  & 0.0161  & 0.0233  & 0.0236  & 0.0343  \\
\midrule
\multirow{5}[2]{*}{On-device Rec} & LLRec & 0.1607  & 0.2115  & 0.1893  & 0.2829  & 0.0054  & 0.0095  & 0.0080  & 0.0163  \\
      & RARec-D & 0.1688  & 0.2316  & 0.2067  & 0.3228  & 0.0075  & 0.0132  & 0.0124  & 0.0227  \\
      & TransRec-D & 0.1853  & 0.2524  & 0.2234  & 0.3403  & 0.0084  & 0.0140  & 0.0128  & 0.0248  \\
      & DCCL  & 0.1629  & 0.2210  & 0.1943  & 0.3245  & 0.0068  & 0.0124  & 0.0116  & 0.0222  \\
      & PREFER & 0.1762  & 0.2303  & 0.1966  & 0.3161  & 0.0059  & 0.0096  & 0.0088  & 0.0179  \\
\midrule
\multirow{4}[2]{*}{Device-Cloud Rec} & DUET  & 0.2020  & 0.2739  & 0.2345  & 0.3768  & 0.0105  & 0.0157  & 0.0135  & 0.0278  \\
      & LSC4Rec  & 0.2425  & 0.3136  & 0.2669  & 0.4186  & 0.0158  & 0.0237  & 0.0257  & 0.0367  \\
      & CDA4Rec  & 0.2367  & 0.3076  & 0.2749  & 0.4296  & 0.0162  & 0.0234  & 0.0250  & 0.0404  \\
      & SCaLRec & \textbf{0.2552 } & \textbf{0.3519 } & \textbf{0.2845 } & \textbf{0.4561 } & \textbf{0.0180 } & \textbf{0.0250 } & \textbf{0.0268 } & \textbf{0.0411 } \\
\bottomrule
\end{tabular}%

  \label{tab:topk}%
  \vspace{-1em}
\end{table*}%

\subsection{Baselines}
\label{sec:exp_baselines}
We compare SCaLRec against strong baselines under three deployment categories.

\subsubsection{Cloud-side Inference Baselines}
\label{sec:exp_cloud_baselines}
These methods are classic or strong sequential recommenders deployed on the cloud.
They provide strong cloud-side inference performance and serve as reference points for semantic-aware modeling.

\begin{itemize}
\item \textbf{LightGCN}~\cite{he2020lightgcn}: A simplified graph-based collaborative filtering model focusing on neighborhood aggregation.
\item \textbf{SURGE}~\cite{chang2021sequential}: A graph-based sequential framework that performs interest-aware graph reconstruction and reasoning.
\item \textbf{SASRec}~\cite{kang2018self}: It learns both long-term and short-term preferences through a self-attention mechanism.
\item \textbf{RARec}~\cite{yu2024ra}: an LLM-based framework that aligns ID embeddings with LLMs via soft prompts.
\item \textbf{TransRec}~\cite{lin2024bridging}: A novel LLM-enhanced recommendation framework for item ranking and generation with structured grounding mechanism.
\end{itemize}

\subsubsection{Device-side Inference Baselines}
\label{sec:exp_device_baselines}
A natural question is whether one can avoid semantic staleness by moving both semantic and structural modeling fully onto devices.
To address this, we include device-side inference baselines that operate without cloud semantic refresh.

\begin{itemize}
\item \textbf{LLRec}~\cite{wang2020next}: A lightweight on-device sequential recommender distilled from a stronger cloud teacher.

\item \textbf{RARec-D}: One possible solution to mitigate the could semantic staleness is to move the RARec to the device. For the inference stage the latest semantic embedding can be generated accompanied with the latest structural embedding. To accommodate the computation ability of the device side, we use LLAMA-1B as the on-device compact LLM instead of the original full capacity LLaMA-8B on the cloud server.

\item \textbf{TransRec-D}: A compact version of TransRec, whose settings are identical to RARec-D for fair comparison.

\item \textbf{DCCL}~\cite{yao2021device}: A device-cloud collaborative learning framework that enables on-device personalization and refines the cloud model via knowledge distillation.
\item \textbf{PREFER}~\cite{guo2021prefer}: A privacy-preserving federated framework that trains on device and aggregates via edge servers.
\end{itemize}

\subsubsection{Cloud-Device Collaborative Inference Baselines}
\label{sec:exp_cd_baselines}
These methods explicitly model cloud-device collaboration and are the most direct comparisons.

\begin{itemize}
\item \textbf{DUET}~\cite{lv2023duet}: A cloud-device framework that generates device-specific parameters via forward propagation to avoid on-device fine-tuning.
\item \textbf{LSC4Rec}~\cite{lv2025collaboration}: A device-cloud collaborative framework combining LLM-based semantic reasoning with lightweight SRMs for real-time adaptation.
\item \textbf{CDA4Rec}~\cite{long2025cloud}: a cloud–device framework with a cloud LLM for semantic reasoning, and an on-device SLM for privacy-preserving real-time ranking, via plan-based task decomposition and coordination.
\end{itemize}

\begin{figure}[t]
  \centering
  \includegraphics[width=\columnwidth]{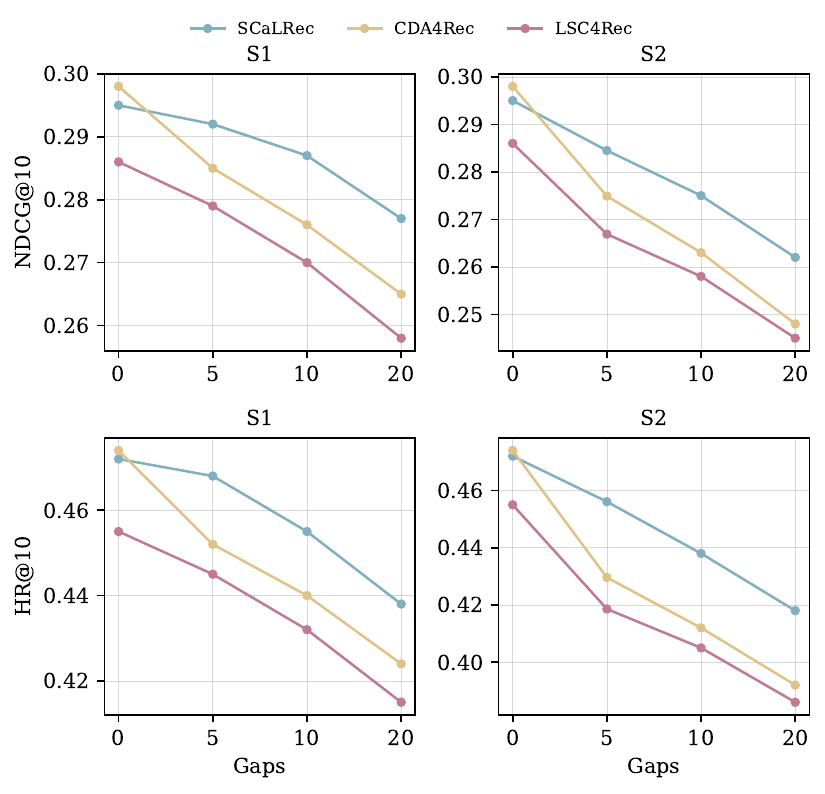}
  \vspace{-1em}
  \caption{Performance under increasing staleness gap $g$.
  We compare the SCaLRec under S1 (fixed-candidate reranking) and S2 (end-to-end cached inference).
  Metrics are NDCG@10 and HR@10 on Foursquare dataset.}
  \label{fig:staleness_curve}
    \vspace{-5pt}

\end{figure}

\begin{figure}[t]
  \centering
  \includegraphics[width=\columnwidth]{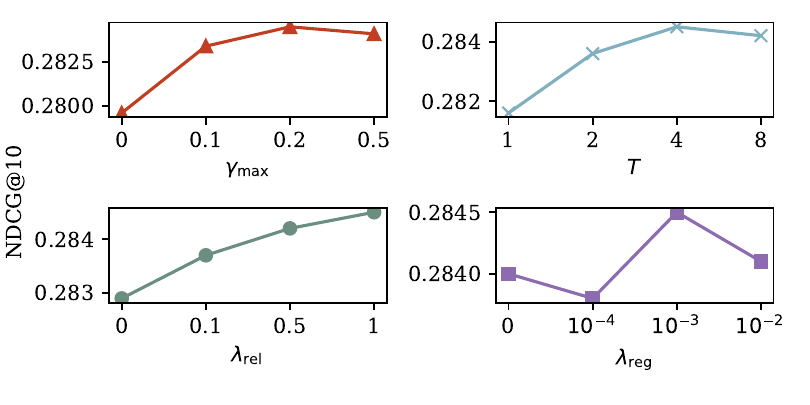}
    \vspace{-3em}

  \caption{Sensitivity of SCaLRec to correction radius, distillation temperature, and loss weights on Foursquare. Performance (NDCG@10) are reported under S2 settings and averaged over different gaps in $\{5,10,20\}$.}
  \label{fig:hyper}
  \vspace{-10pt}
\end{figure}

\subsection{RQ1: Overall Performance}
\label{sec:rq1}
To validate the effectiveness of our proposed method, we compare it with various kinds of baselines. To simulate the semantic staleness needed for cloud recommendation and cloud–device recommendation, we both set the gap equal to 5. This means that the cloud can only see the previous interactions except the latest five interactions. For on-device recommendation, since the inference is only performed locally, whether it is a federated recommendation or a small recommendation performed on-device, we assume that the model can see the latest interactions. Since it is on the device side, there are no privacy or latency concerns. Table~\ref{tab:topk}  presents the recommendation performance results, and we can observe the following key insights.

Across both datasets, SCaLRec achieves the strongest results among all compared methods under the same staleness gap.
The gains are consistent for both NDCG and HR.
Compared with CDA4Rec and LSC4Rec, SCaLRec improves accuracy in this cached setting without changing the evaluation protocol.
This supports our core claim that explicit semantic calibration can add value on top of an established cloud-device backbone.

A second observation is that device-cloud collaboration remains the most competitive regime under realistic constraints.
Pure cloud-side recommenders perform reasonably well when strong semantic representations are available, but they do not exploit the freshest interaction evidence on the device at inference time.
Pure on-device approaches, including federated or lightweight local models, lag behind in accuracy even though they can access the latest interactions.
It suggests that combining a cached cloud semantic signal with a fresh on-device structure signal is still the best trade-off in practice. Among cloud-side baselines, methods that align item semantics with pretrained language models, such as RARec and TransRec, are generally stronger than classic sequential or graph models.

Finally, simply pushing LLM-style modeling onto the device does not close the gap.
The lightweight on-device variants of RARec and TransRec show a noticeable drop compared with their cloud versions.
Other on-device baselines show a similar limitation.
This makes the cached cloud semantic signal a realistic component in a deployed pipeline, and it also motivates why we focus on mitigating stale semantics instead of removing the cloud semantic channel entirely.

\subsection{RQ2: Performance Degradation under Increasing Staleness}
\label{sec:rq2}

To investigate how different gap sizes influence the proposed method, the gap intervals are systematically selected from {0, 5, 10, 20}. The proposed method, along with two popular cloud–device recommendation models, CDA4Rec and LSC4Rec, is evaluated on the Foursquare dataset, with recommendation performance reported in terms of NDCG@10 and HR@10. To examine the phenomenon of semantic staleness in more detail, we consider two gap settings.
\textbf{S1} isolates semantic mismatch by keeping the candidate set fixed and only replacing the semantic embedding with an earlier cached version.
\textbf{S2} reflects an end-to-end cached pipeline, in which both the semantic embedding and the candidate set come from the same stale cloud snapshot.
From Figure~\ref{fig:staleness_curve}, we can draw the following insights and conclusions:

\textbf{Monotonic degradation with increasing gap.}
All methods show a clear downward trend as the staleness gap grows, under both S1 and S2.
The drop is steady rather than erratic, suggesting the effect is systematic once semantics are reused across requests.

\textbf{S2 is consistently harder than S1.}
The curves under S2 fall faster than those under S1.
This is expected since S2 compounds two sources of staleness: the semantic signal and the retrieval snapshot that produces the candidate pool.


\textbf{Widening margins at larger gaps.}
As the gap increases, SCaLRec separates from the backbones and maintains a larger margin at the tail of the range.
The separation is visible in both settings, and it is more pronounced in S2, where stale retrieval makes the reranker more sensitive to semantic mismatch.

\subsection{RQ3: Hyperparameter Sensitivity}
\label{sec:rq3}

We study how SCaLRec behaves when we change a few key hyperparameters.
We run this analysis on Foursquare.
We report NDCG@10 only.
We evaluate under S2 and average results over gaps $\mathcal{G}=\{5,10,20\}$.
For each sweep, we vary one hyperparameter and keep the others fixed. Results are shown in Figure~\ref{fig:hyper}.

\textbf{Correction radius $\gamma_{\max}$.}
$\gamma_{\max}$ controls how far the calibrator is allowed to move the cached semantic embedding.
When $\gamma_{\max}=0$, the model reduces to the backbone without semantic correction, and performance drops.
Increasing $\gamma_{\max}$ from small to moderate values improves performance.
The improvement then saturates.
Very large radii do not bring further gains, and can slightly hurt due to occasional over-correction under mild mismatch.

\textbf{Distillation temperature $T$.}
$T$ affects how soft the teacher distribution is on the cached candidate set.
With small $T$, the supervision becomes sharp and the student can overfit to top-ranked candidates under noisy stale snapshots.
A moderate $T$ is more stable across gaps.
Very large $T$ can weaken the supervision signal, and the calibration becomes less targeted.
Overall, we observe a peak around a moderate temperature, with mild degradation on both sides.

\textbf{Reliability loss weight $\lambda_{\mathrm{rel}}$.}
$\lambda_{\mathrm{rel}}$ controls how strongly we train the reliability estimator.
If $\lambda_{\mathrm{rel}}=0$, the estimator is effectively removed.
The calibrator then lacks a gate on correction strength, so it may perturb semantics even when the snapshot is already aligned.
Introducing a non-zero $\lambda_{\mathrm{rel}}$ stabilizes training and improves performance.
If $\lambda_{\mathrm{rel}}$ becomes too large, the optimization over-emphasizes fitting the reliability target, and the gain from the calibrator becomes less noticeable.

\textbf{Regularization weight $\lambda_{\mathrm{reg}}$.}
$\lambda_{\mathrm{reg}}$ penalizes large proposal vectors from the calibrator.
Without regularization, the calibrator can become slightly aggressive, which is not always helpful under small gaps.
A small amount of regularization improves robustness and reduces jitter.
If $\lambda_{\mathrm{reg}}$ is too large, the calibrator under-corrects when mismatch is strong, and performance drops slightly.

\begin{figure}[t]
  \centering
  \includegraphics[width=0.8\columnwidth]{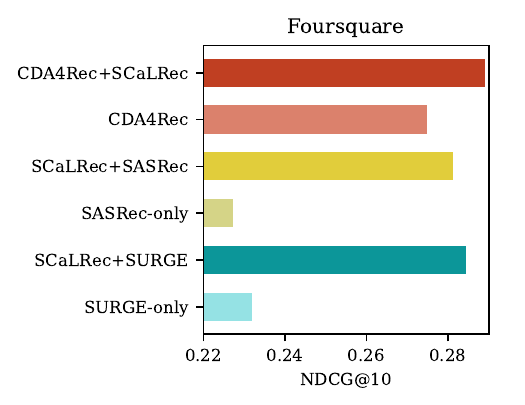}
    \vspace{-0.5em}

  \caption{Model-agnostic integration on Foursquare under semantic staleness. We plug SCaLRec into different on-device structural backbones (SASRec and SURGE) and into a cloud-device framework (CDA4Rec).}
  \label{fig:agnostic}
  \vspace{-1em}
\end{figure}

\begin{figure}[t]
  \centering
  \includegraphics[width=\columnwidth]{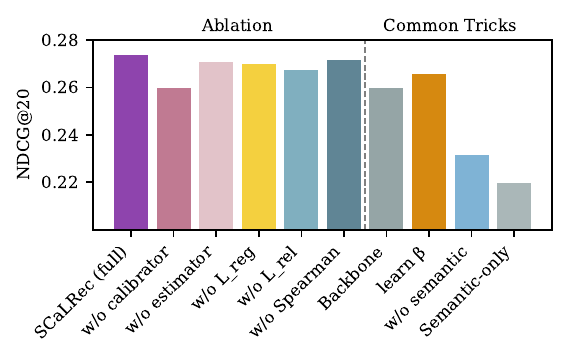}
    \vspace{-1em}

\caption{Ablation and common-trick on Foursquare under S2 (cached pipeline), reporting NDCG@10 averaged over $g\in\{5,10,20\}$.}
\vspace{-1em}
  \label{fig:ablation}
\end{figure}

\subsection{RQ4: Model-Agnostic Plug-in Evaluation}
\label{sec:rq4}
In this section, we investigate whether the proposed method could be integrated with other classical methods seamlessly. There are two aspects to evaluate the model agnosticism. First, whether the proposed method can be integrated with other cloud-device methods (e.g., CDA4Rec), helping them mitigate the semantic staleness effect. Second, within the semantic calibration of SCaLRec, there needs to be a model on the device side to learn the structural signal through users’ recent interactions. Is SCaLRec only feasible for one particular method, or does it have general applicability to different classical methods (e.g., SURGE or SASRec)? Figure~\ref{fig:agnostic} shows the recommendation performance on the FourSquare dataset.

We first test whether SCaLRec can be used as a plug-in module on top of different on-device structural backbones. Adding SCaLRec on top of SURGE improves NDCG@10 compared with \textit{SURGE-only}. A similar pattern holds for SASRec. The improvement is visible even though the two backbones model interaction sequences in different ways. This suggests that the calibrator does not rely on a specific architecture. It mainly requires a short-term structural signal that is computed from recent interactions, and it can use this signal to compensate the stale semantic channel under the same reranking interface.

We then examine the integration with an existing cloud-device framework. CDA4Rec already benefits from LLM-enabled semantic user modeling on the cloud and a context-aware strategy planner that decides how cloud and device components should collaborate under different user scenarios. When we further attach SCaLRec to CDA4Rec, the combined system achieves the best performance among all compared settings. The result indicates that semantic calibration is complementary to strategy planning. Planning selects an appropriate collaboration path when cloud interactions are available, but it does not remove the mismatch that arises once cloud semantics are cached and reused. Calibration targets this mismatch directly at inference time. In practice, this combination provides a stronger and more robust reranking behavior under intermittent cloud availability.

\subsection{RQ5: Ablations and Common Tricks}
\label{sec:rq5}

We study how each design choice contributes to the final performance.
Figure~\ref{fig:ablation} reports the results on Foursquare under the cached pipeline (S2), averaged over gaps $g\in\{5,10,20\}$.

\textbf{Ablation on core modules and losses.}
Removing the calibrator collapses the method back to the cached fusion backbone.
This drop is large, so the gain does not come from tuning or randomness.
Removing the estimator gives a smaller drop.
The calibrator still helps in this case, but it becomes easier to over-correct at some gaps.
The loss terms show a similar pattern.
Dropping $\mathcal{L}_{\mathrm{rel}}$ hurts more than dropping $\mathcal{L}_{\mathrm{reg}}$.
This matches the intuition that the method needs a meaningful reliability target, not only a smoothness prior.
Learning a dynamic fusion weight $\beta$ is a reasonable baseline.
It improves over the backbone, but it still trails SCaLRec.

\textbf{Common tricks and lower-bound baselines.}
We remove the Spearman agreement cue $\omega_{u,t}$ from $c_{u,t}$.
The performance drops, but it does not crash.
So the cue is useful, but the method does not depend on a single feature.
We also test two simple fallbacks.
Semantic-only with cached semantics is weak.
Structure-only on the cached candidate set is stronger, but it remains far below calibrated fusion.
In this setup, discarding the semantic channel is not a good default.
It is the kind of hack that looks safe but loses too much.

Learning $\beta$ is a naive solution for balancing information from two channels. Therefore, we implement the same calibrator architecture as the gating method, which is also trained offline on the device side with the same calibrator inputs. It acts as a pure gating mechanism rather than the proposed method, which actively adjusts the semantic channel by recovering valuable information. Learning $\beta$ slightly outperforms the case where no action is taken to address cloud semantic staleness, but it still lags behind the proposed method. This reveals the effectiveness of the estimation and calibration procedures.

\section{Conclusion}

In this paper, we study semantic staleness in LLM-enabled device–cloud sequential reranking, where a cloud semantic user embedding is often cached and reused under intermittent refresh. This cached semantics can become misaligned with the user’s latest interactions on device, leading to consistent reranking degradation under controlled staleness gaps.

We propose ScaLRec, which keeps the standard dot-product reranking interface and improves the use of cached semantics with two lightweight on-device modules. A reliability estimator predicts when cached semantics is risky, and a semantic calibrator corrects the cached embedding instead of simply down-weighting it. The calibrator is trained offline via knowledge distillation to match the ranking behavior induced by fresh semantics, but it runs fully on device at inference time. Experiments across two datasets and staleness levels show consistent gains over strong baselines. ScaLRec does not address when to refresh cloud semantics, and it focuses on correcting cached user semantics.


\bibliographystyle{ACM-Reference-Format}
\bibliography{main}

\appendix

\end{document}